\documentclass{article}
\usepackage{bm, graphicx, amsmath}
\title{Finding more than one path through a simple maze with a quantum walk}
\author{Mark Hillery \\
Department of Physics \\  Hunter College of the City University of New York, \\ 695 Park Avenue, New York, NY 10065 USA \\ and Physics Program \\ Graduate Center of the City University of New York \\ 365 Fifth Avenue, New York, NY 10016}

\begin{document}
\maketitle

\begin{abstract}
We study quantum walks through chains consisting of two and three star graphs.  The first star has a distinguished vertex labelled START and the last has one labelled END.  There are multiple paths between these two vertices, and the object is to find these paths.  We show that a quantum walk can do this with a quantum speedup.
\end{abstract}

\section{Introduction}
Quantum walks on graphs have proven to be a useful framework in which to explore quantum algorithms \cite{walk1,walk2,walk3,walk4,walk5}.  One of the most common kind of algorithm studied in this context is one for a search problem.  In its simplest form, one of the vertices of the graph behaves differently than the others, and the object is to find that vertex.  This has been done for a number of highly symmetric graphs, such as the hypercube \cite{shenvi,potocek}, grids in different dimensions \cite{grid1,grid2}, and the complete graph \cite{grid2,complete2}.  The role of the symmetry of the graph in a quantum-walk search has been explored \cite{wong}, and it has been shown that a quantum-walk search is optimal for almost all graphs \cite{Omar}.  The initial state of the walk cannot incorporate any knowledge of the distinguished vertex, and it is usually an equal superposition of all vertices, in the case of a coined walk, or an equal superposition of all edges, in the case of a scattering walk.  These walks typically achieve a quadratic speedup over what is possible classically, that is, they only require a number of steps that is of the order of the square root of the number of vertices in the graph.  It should be noted that by constructing a  quantum circuit that implements a quantum walk, these graph search problems can be rephrased as searches involving calls to an oracle.  For an explicit example of this see \cite{complete2}.

It is possible to find more elaborate structures than single vertices.  For example, it is possible to find an extra edge that breaks the symmetry of a graph, or where two graphs are connected. \cite{feldman,anomaly1}.  It is also possible to  find a general graph attached to one of the edges of a star graph \cite{cottrell}.  More recently, quantum walks have been used to find paths through graphs.  In \cite{reitzner}, a path through a graph consisting of a line of connected star graphs was found with a quadratic speedup.  The first star had a distinguished vertex labelled START, and the last star had a distinguished vertex labelled END, and there was only one path from START to END through the graph.  This is equivalent to finding the way through a maze, consisting of interconnected rooms, in which each room has many doors, but  only one leads to the next room.  It is also possible to find a path through a tree from the root to a distinguished leaf \cite{koch}.

Here we will again be interested in finding paths through graphs from START to END through a graph  consisting of connected stars, but now in the case that there is more than one path.  We will look at two cases.  We will begin with just two stars with two paths from START to END.  In this case, there are few surprises, and after a number of steps of order of the square root of the number of vertices, the particle becomes localized on the two paths.  We then add a third star between the first two, and allow an arbitrary number of paths.  Here new features emerge.

Finally, it should be noted that quantum walks are not just theoretical constructs.  They have been realized in the laboratory using a number of different systems \cite{PeLaPoSoMoSi08} - \cite{schreiber}.

\section{Two stars}
In this paper we will be using the scattering quantum walk in which the particle ``scatters'' off the vertices of the graph \cite{walk4,complete2}.  The particle making the walk sits on the edges of the graph instead of the vertices.  Each edge has two orthogonal states.  If the edge connects vertices $j$ and $k$, then one state is $|j,k\rangle$ corresponding to the particle going from $j$ to $k$, and the other is $|k,j\rangle$ corresponding to the particle going from $k$ to $j$.  The collection of all of these states, two for each edge, forms an orthonormal basis for the Hilbert space in which the walk takes place.  In addition to the Hilbert space we need a unitary operator that advances the walk one step.  In the scattering walk each vertex acts as a scattering center and is described by a local unitary operator that maps states entering the vertex to states leaving the vertex.  The unitary operator that advances the walk one step, $U$, is simply made up of the action of all of the local unitary operators at the vertices. For a vertex, $j$, with $n > 2$ edges connected to it, we will use the operator
\begin{equation}
U|k,j\rangle = -r |j,k\rangle + t \sum_{l=1,l\neq k}^{n} |j,l\rangle ,
\end{equation}
where $r=(n-2)/n$, is the amplitude to be reflected, and $t=2/n$, is the amplitude to be transmitted.  This type of vertex behaves in the same way no matter from which edge it is entered.  For the case $n=2$, we will assume the particle is transmitted with no amplitude for reflection.

For our first graph, we have two stars, with centers $A_{1}$ and $A_{2}$,  and each has $N$ edges emanating from the center, and we will assume that $N$ is large.  The outer vertices on the first star are $B_{11}$ through $B_{1N}$.  The vertices $B_{12}$ and $B_{13}$ are common to both stars, so the outer vertices on the second star are $B_{21}$, $B_{12}$, $B_{13}$, and $B_{24}$ to $B_{2N}$.  The vertex $B_{11}$ is the START vertex, and $B_{21}$ is the END vertex.  These two vertices reflect with a phase factor of $-1$.  The vertices $B_{12}$ and $B_{13}$ transmit with no reflection, and all the other outer vertices reflect with a phase factor of $1$.  This means there are two paths from START to END, one via $B_{12}$ and one via $B_{13}$.

It is possible to simplify the problem by taking advantage of the symmetry of the problem and defining new states that are collections of edges \cite{krovi,complete2}.  We define the following states:
\begin{eqnarray}
|\psi_{1}\rangle & = & \frac{1}{\sqrt{2(N-3)}} \left( \sum_{j=4}^{N} |A_{1},B_{1j}\rangle -\sum_{j=4}^{N} |A_{2},B_{2j}\rangle \right) \nonumber \\
|\psi_{2}\rangle & = &  \frac{1}{\sqrt{2(N-3)}} \left( \sum_{j=4}^{N} |B_{1j}, A_{1}\rangle -\sum_{j=4}^{N} |B_{2j},A_{2}\rangle \right) \nonumber \\ 
|\psi_{3}\rangle & = & \frac{1}{\sqrt{6}}( |B_{11},A_{1}\rangle + |B_{12},A_{1}\rangle + |B_{13},A_{1}\rangle \nonumber \\
& & - |B_{21},A_{2}\rangle - |B_{12},A_{2}\rangle - |B_{13}, A_{2}\rangle ) \nonumber \\
|\psi_{4}\rangle & = & \frac{1}{\sqrt{6}} ( |A_{1},B_{11}\rangle + |A_{1},B_{12}\rangle + |A_{1},B_{13}\rangle \nonumber \\
& & - |A_{2},B_{21}\rangle - |A_{2},B_{12}\rangle - |A_{2},B_{13}\rangle ) .
\end{eqnarray}
Note that the states $|\psi_{3}\rangle$ and $|\psi_{4}\rangle$ are superpositions of the states on the paths between START and END.  The action of the unitary operator that advances the walk one step is
\begin{eqnarray}
U|\psi_{1}\rangle & = & |\psi_{2}\rangle \nonumber \\
U|\psi_{2}\rangle & = & (r-2t) |\psi_{1}\rangle + t\sqrt{3(N-3)} |\psi_{4}\rangle \nonumber \\
U|\psi_{3}\rangle & = & -(r-2t) |\psi_{4}\rangle + t\sqrt{3(N-3)} |\psi_{1}\rangle \nonumber \\
U|\psi_{4}\rangle & = & - |\psi_{3}\rangle .
\end{eqnarray}
The graph is bipartite, so we can reduce the dimensionality of the problem from four to two by looking at the action of $U^{2}$,
\begin{eqnarray}
U^{2}|\psi_{2}\rangle & = & (r-2t) |\psi_{2}\rangle - t\sqrt{3(N-3)} |\psi_{3}\rangle \nonumber \\
U^{2}|\psi_{3}\rangle & = & (r-2t)|\psi_{3}\rangle  + t\sqrt{3(N-3)} |\psi_{2}\rangle .
\end{eqnarray}
This transformation reduces to the $2\times 2$ matrix
\begin{equation}
\left( \begin{array}{cc} r-2t & t\sqrt{3(N-3)} \\ -t\sqrt{3(N-3)}  & r-2t \end{array} \right) ,
\end{equation}
which has eigenvalues $\lambda = r-2t \pm it\sqrt{3(N-3)} $.  These eigenvalues can be approximated up to order $1/\sqrt{N}$ by $\exp (\pm i\gamma )$, where $\gamma = t\sqrt{3(N-3)}$.  The eigenstates are $|u_{\pm}\rangle = (1/\sqrt{2}) (|\psi_{2}\rangle \pm i |\psi_{3}\rangle )$.  For the initial state we choose
\begin{equation}
|\psi_{in}\rangle = \sqrt{\frac{N-3}{N}} |\psi_{2}\rangle + \sqrt{\frac{3}{N}} |\psi_{3}\rangle ,
\end{equation}
which is an equal superposition of outgoing states on the first star minus outgoing states on the second star.  After $2n$ steps this state becomes, approximately,
\begin{equation}
U^{2n}|\psi_{in}\rangle = \cos (n\gamma ) |\psi_{2}\rangle -i\sin (n\gamma) |\psi_{3}\rangle .
\end{equation} 
When $n\gamma  =\pi /2$, the particle is in the state $|\psi_{3}\rangle$, which is localized on both paths connecting START and END.  Therefore, by repeating the walk several times and measuring the location of the particle after $2n=\pi/(\gamma ) = O(\sqrt{N})$ steps, we can find both paths.

\section{Three stars}
We would now like to move on to a more complicated situation.  We will insert an additional star between the two we had in the previous section and allow there to be more than two paths.  This will result in some new features.

We now have three stars, which, going from left to right, we shall denote as $1$, $2$ and $3$ (see Fig.\ 1).  Star $1$ has the vertex START and star $3$ has the vertex END.  As before, each star has $N$ prongs, and star $1$ and $2$ share $m-1$ vertices and stars $2$ and $3$ share $m-1$ vertices, where $m\ll N$.   The case $m=2$ corresponds to just a single path connecting START and END.  The central vertices of the stars are denoted by $A_{1}$, $A_{2}$, and $A_{3}$. As before, the outer vertices are denoted by $B$ and corresponding subscripts.  The START vertex is $B_{11}$ and END is $B_{31}$.  Vertices $B_{12}$ to $B_{1m}$ are shared by stars one and two, and each of these vertices is connected to two edges, one going to $A_{1}$ and one going to $A_{2}$.  Similarly, vertices $B_{32}$ to $B_{3m}$ are shared by stars $2$ and $3$, with each vertex connected to two edges, one going to $A_{2}$ and one going to $A_{3}$.  The remaining vertices, $B_{1,m+1}$ to $B_{1N}$ are connected to $A_{1}$, $B_{21}$ to $B_{2,N-2m+2}$ are connected to $A_{2}$, and $B_{3,m+1}$ to $B_{3N}$ are connected to $A_{3}$.  All outer vertices reflect the particle with a phase factor of $1$ except for $B_{11}$ and $B_{31}$, which reflect it with a phase factor of $-1$.  The vertices connected to two edges simply transmit the particle with no reflection.

\begin{figure}
\begin{center}
\includegraphics[width=8.2cm]{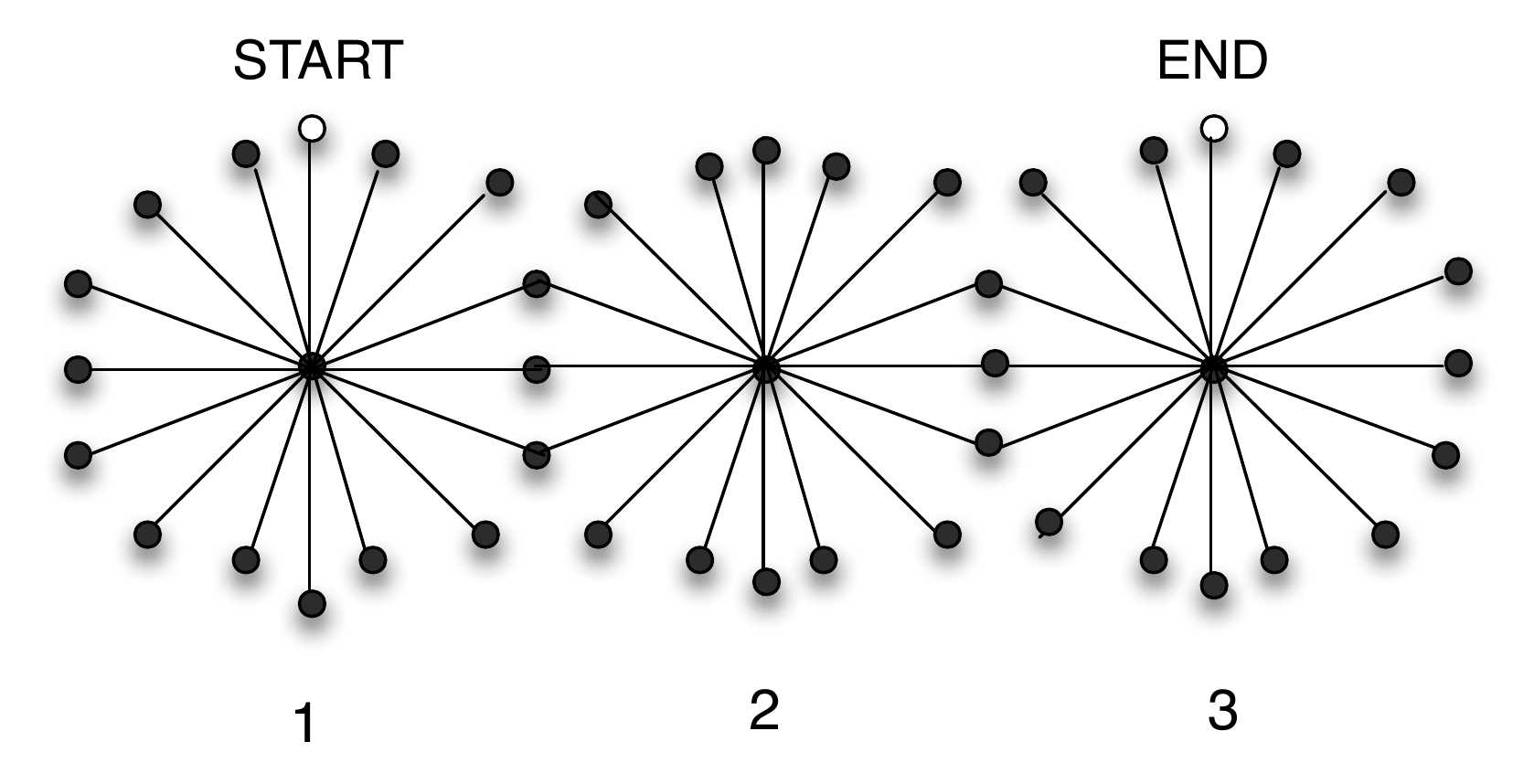}
\end{center}
\caption{A chain of three stars with multiple paths from START to END.}
\label{fig:stars}
\end{figure}

In order to discuss the dynamics, it is useful to group the states.  Define, for star $1$,
\begin{eqnarray}
|A_{1},B_{1s}\rangle & = & \frac{1}{\sqrt{N-m}} \sum_{j=m+1}^{N} |A_{1},B_{1j}\rangle \nonumber \\
|A_{1},B_{1c}\rangle & = & \frac{1}{\sqrt{m-1}} \sum_{j=2}^{m} |A_{1},B_{1j}\rangle  ,
\end{eqnarray}
with $|B_{1s},A_{1}\rangle$ and $|B_{1c},A_{1}\rangle$ being defined similarly, just with the positions of $A$ and $B$ being interchanged in the above equations.  Note that $|A_{1},B_{1s}\rangle$ contains only edges that are confined to star $1$, while $|A_{1},B_{1c}\rangle$ contains edges in star $1$ that are connected to star $2$.  For star $2$ we have the states
\begin{eqnarray}
|B_{2cl},A_{2}\rangle  & = & \frac{1}{\sqrt{m-1}} \sum_{j=2}^{m} |B_{1j},A_{2}\rangle \nonumber \\
|B_{2s},A_{2}\rangle & = & \frac{1}{\sqrt{N-2m+2}} \sum_{j=1}^{N-2m+2} |B_{2j},A_{2}\rangle \nonumber \\
|B_{2cr},A_{2}\rangle & = & \frac{1}{\sqrt{m-1}} \sum_{j=2}^{m} |B_{3j},A_{2}\rangle ,
\end{eqnarray}
and the corresponding reversed states with $A$ and $B$ interchanged.  Here, $|B_{2cl},A_{2}\rangle$ contains edges that are ccnnected to star $1$, $|B_{2s},A_{2}\rangle$ contains edges that are confined to star $2$, and $|B_{2cr},A_{2}\rangle$ contains edges that are connected to star $3$.  Finally, for star $3$ we have
\begin{eqnarray}
|A_{3},B_{3s}\rangle & = & \frac{1}{\sqrt{N-m}} \sum_{j=m+1}^{N} |A_{3},B_{3j}\rangle \nonumber \\
|A_{3},B_{3c}\rangle & = & \frac{1}{\sqrt{m-1}} \sum_{j=2}^{m} |A_{3},B_{3j}\rangle  ,
\end{eqnarray}
with the corresponding reversed states.  The action of $U$, the operator that advances the walk one step, on the outgoing states is simple, for example, for star $1$,
\begin{eqnarray}
U|A_{1},B_{11}\rangle & = & -|B_{11},A_{1}\rangle \nonumber \\
U|A_{1},B_{1s}\rangle & = & |B_{1s},A_{1}\rangle \nonumber \\
U|A_{1},B_{1c}\rangle & = & |B_{1c},A_{2}\rangle ,
\end{eqnarray}
and similarly for the other two stars.  The situation for the ingoing states is more complicated.  For star $1$ we have
\begin{eqnarray}
U|B_{11},A_{1}\rangle & = & -r|A_{1},B_{11}\rangle + t\sqrt{N-m} |A_{1},B_{1s}\rangle + t\sqrt{m-1} |A_{1},B_{1c}\rangle \nonumber \\
U|B_{1c},A_{1}\rangle & = & [-r+t(m-2)] |A_{1},B_{1c}\rangle + t\sqrt{m-1} |A_{1},B_{11}\rangle \nonumber \\
& & + t\sqrt{(m-1)(N-m)} |A_{1},B_{1s}\rangle \nonumber \\
U|B_{1s},A_{1}\rangle & = & (1-mt) |A_{1},B_{1s}\rangle + t\sqrt{N-m} |A_{1},B_{11}\rangle \nonumber \\
& & + t\sqrt{(m-1)(N-m)} |A_{1},B_{1c}\rangle .
\end{eqnarray}
For star $2$
\begin{eqnarray}
U|B_{2cl},A_{2}\rangle & = & -[1-t(m-1)] |A_{2},B_{2cl}\rangle + t(m-1)|A_{2},B_{2cr}\rangle 
\nonumber \\
& & + t\sqrt{(m-1)(N-2m+2)} |A_{2},B_{2s}\rangle \nonumber \\
U|B_{2s},A_{2}\rangle & = & [1-2t(m-1)] |A_{2},B_{2s}\rangle + t\sqrt{(m-1)(N-2m+2)} \nonumber \\
& & (|A_{2},B_{2cl}\rangle + |A_{2}, B_{2cr}\rangle ) ,
\end{eqnarray}
with $U|B_{2cr},A_{2}\rangle$ being easily found from the expression for $U|B_{2cl},A_{2}\rangle$.  The expressions for star $3$ are similar to those for star $1$.

The problem can be simplified by noting two things.  First, the graph is bipartite, so by considering the action of $U^{2}$ instead of $U$, we can eliminate half the states.  Second, we can take advantage of the symmetry between stars $1$ and $3$ to combine their states.  We now define the five orthonormal states
\begin{eqnarray}
|\psi_{1}\rangle & = & \frac{1}{\sqrt{2}} (B_{1s},A_{1}\rangle + |B_{3s},A_{3}\rangle ) \nonumber \\
|\psi_{2}\rangle & = & \frac{1}{\sqrt{2}} (|B_{11},A_{1}\rangle + |B_{31},A_{3}\rangle ) \nonumber \\
|\psi_{3}\rangle & = & \frac{1}{\sqrt{2}} (|B_{1c},A_{1}\rangle + |B_{3c},A_{3}\rangle ) \nonumber \\
|\psi_{4}\rangle  & = & \frac{1}{\sqrt{2}} (|B_{2cl},A_{2}\rangle + |B_{c2r},A_{2}\rangle ) \nonumber \\
\psi_{5}\rangle  & = & |B_{2s},A_{2}\rangle .
\end{eqnarray}
In this basis, the operator $U^{2}$ can be expressed as the matrix
\begin{equation}
M = \left( \begin{array}{ccccc} 1-mt & t\sqrt{N-m} & t\mu & 0 & 0  \\ 
-t\sqrt{N-m} & r & -t\sqrt{m-1} & 0 & 0 \\    0 & 0 & 0 & 1-2(m-1)t & t\nu \\
t\mu & t\sqrt{m-1} & -1 + t(m-1) & 0 & 0 \\ 0 & 0 & 0 & t\nu & 1 - 2(m-1)t \end{array} \right) ,
\end{equation}
where $\mu = \sqrt{(m-1)(N-m)}$ and $\nu = \sqrt{2(m-1)(N-2m+2)}$.

We now need to find the eigenvalues and eigenvectors of this matrix. The characteristic equation is
\begin{eqnarray}
\lambda^{5} -  [3(1-mt) + t] \lambda^{4} + [2-(3m-1)t+4(m-1)t^{2}] \lambda^{3} \nonumber \\
+ [2-(3m-1)t+4(m-1)t^{2}]  \lambda^{2} -[3-(3m-1)t] \lambda +1 = 0.
\end{eqnarray}
One of the roots of this equation is $\lambda = -1$.  Factoring out $\lambda + 1$, we obtain a quartic equation, which can be factored, resulting in two quadratic equations.  For the remaing eigenvalues of $M$ we find
\begin{eqnarray}
\lambda & = & e^{\pm i\sqrt{\gamma_{+}t/2}} + O(1/N) \nonumber \\
\lambda & = & e^{\pm i\sqrt{\gamma_{-}t/2}} + O(1/N)  ,
\end{eqnarray}
where
\begin{equation}
\gamma_{\pm} = (3m-1) \pm \sqrt{(3m-1)^{2} -16(m-1)} .
\end{equation}
We shall denote the eigenvectors by $|u_{jk}\rangle$, where $j=\pm$ and $k=\pm$.  The index $j$ indicates whether the eigenvector is for $\gamma_{+}$ or $\gamma_{-}$, and the index $k$ indicates whether it is for $\exp (+i\sqrt{\gamma_{j}t/2})$ or $\exp (-i\sqrt{\gamma_{j}t/2})$.  Denoting the components of the eigenvectors by $w_{\pm} x_{j}$, where $j=1,2,\ldots 5$ and $w_{\pm}$ is a normalization factor, we find, to lowest order in $1/N$, for the components of the $\gamma_{+}$ eigenvector $|u_{++}\rangle$
\begin{eqnarray}
x_{1}=1 & x_{2} = \frac{2i}{\sqrt{\gamma_{+}}} & x_{3} = \frac{i(\gamma_{+}-4)}{2\sqrt{\gamma_{+}(m-1)}} \nonumber \\
x_{4}=-x_{3} & x_{5} = - \frac{\sqrt{2}}{\gamma_{+}} (\gamma_{+} - 4) & ,
\end{eqnarray} 
and
\begin{equation}
w_{+}^{2}=\frac{2(3m-1)\gamma_{+}-16(m-1)}{4(9m-11)\gamma_{+}-32(3m-5)} .
\end{equation}
For $|u_{+-}\rangle$ we have 
\begin{eqnarray}
x_{1}=1 & x_{2} = \frac{-2i}{\sqrt{\gamma_{+}}} & x_{3} = \frac{-i(\gamma_{+}-4)}{2\sqrt{\gamma_{+}(m-1)}} \nonumber \\
x_{4}=-x_{3} & x_{5} = - \frac{\sqrt{2}}{\gamma_{+}} (\gamma_{+} - 4) & .
\end{eqnarray} 
The components for the eigenvectors corresponding to $\gamma_{-}$ are given by the above expressions with $\gamma_{+}$ replaced by $\gamma_{-}$, and similarly $w_{-}$ is given by the expression for $w_{+}$ with $\gamma_{+}$ replaced by $\gamma_{-}$.

For the initial state of the system, we will take an equal superposition of all ingoing edges, but the amplitudes on stars $1$ and $3$ are positive and those on star $2$ are negative.  This state can be expressed as 
\begin{eqnarray}
|\psi_{in}\rangle & = & \sqrt{\frac{2}{3}} |\psi_{1}\rangle - \sqrt{\frac{1}{3}} |\psi_{5}\rangle + O(N^{-1/2}) \nonumber \\
& = & w_{+} \left[ \sqrt{\frac{2}{3}} + \sqrt{\frac{1}{3}} \frac{\sqrt{2}}{\gamma_{+}} (\gamma_{+}-4) \right](|u_{++}\rangle + |u_{+-}\rangle ) \nonumber \\
 & & +w_{-} \left[ \sqrt{\frac{2}{3}} + \sqrt{\frac{1}{3}} \frac{\sqrt{2}}{\gamma_{-}} (\gamma_{-}-4) \right](|u_{--}\rangle + |u_{-+}\rangle ) .
\end{eqnarray}
After $2n$ steps, the state of the walking particle is 
\begin{eqnarray}
|\psi_{n}\rangle & = & w_{+} \left[ \sqrt{\frac{2}{3}} + \sqrt{\frac{1}{3}} \frac{\sqrt{2}}{\gamma_{+}} (2\gamma_{+}-4) \right] 
(e^{in\sqrt{\gamma_{+}t/2}}|u_{++}\rangle + e^{-in\sqrt{\gamma_{+}t/2}}|u_{+-}\rangle ) \nonumber \\
 & & +w_{-} \left[ \sqrt{\frac{2}{3}} + \sqrt{\frac{1}{3}} \frac{\sqrt{2}}{\gamma_{-}} (2\gamma_{-}-4) \right] 
 (e^{-in\sqrt{\gamma_{-}t/2}}|u_{--}\rangle +e^{in\sqrt{\gamma_{-}t/2}} |u_{-+}\rangle ) .
 \nonumber \\ 
\end{eqnarray}
The states $|u_{++}\rangle - |u_{+-}\rangle$ and $|u_{--}\rangle - |u_{-+}\rangle$ are localized on the paths from START to END.  However, since $\gamma_{+} \neq \gamma_{-}$, the two terms in the above equation will not be localized on the paths after the same number of steps.  An exception is the case $m=2$ (one path), for which the $\gamma_{-}$ term vanishes.  However, the probability to be in the normalized states $w_{+}|u_{++}\rangle$ or $w_{+}|u_{+-}\rangle$, which we shall call $p_{+}$,  is considerably larger than that to be in $w_{-}|u_{-+}\rangle$ or $w_{-}|u_{--}\rangle$.  For $m=3$, $p_{+}$ is $0.97$, for $m=6$ it is $0.93$, and for large $m$ it goes to $0.89$.  Therefore, the terms corresponding to $\gamma_{+}$ dominate the behavior of the state.  That means that we can measure the position of the particle after $2n_{0}=2\pi /\sqrt{2\gamma_{+}t} = O(\sqrt{N})$ steps, and with high probability we will find an edge lying on one of the paths from START to END.  In the case $m=3$ (two paths), that probability is $0.98$. 

\begin{figure}
\begin{center}
\includegraphics[width=8.2cm]{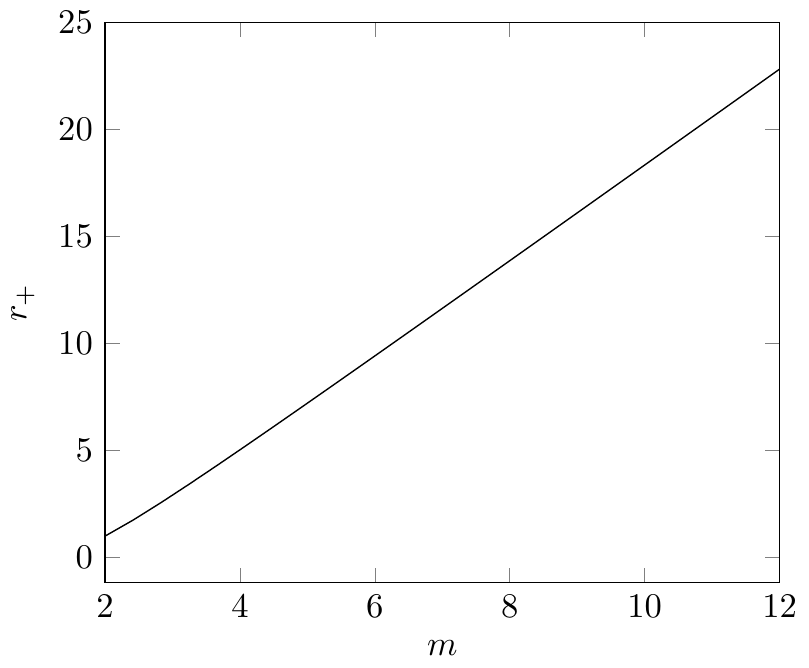}
\end{center}
\caption{The ratio of the probability to be in the state $|\psi_{3}\rangle$ to that to be in the state $|\psi_{2}\rangle$.  }
\label{fig:graph}
\end{figure}

A second issue is that after $2n_{0}$ steps, the probabilities to be in the different states $|\psi_{j}\rangle$, $j=2,3,4$, along the paths are not the same.  If we plot the ratio, $r_{+}$, of the probability to be in the state $|\psi_{3}\rangle$ to that to be in the state $|\psi_{2}\rangle$, we see that it increases roughly linearly with $m$ (see Fig.\ 2).  However, $|\psi_{3}\rangle$ contains roughly $m$ times more individual path states than does $|\psi_{2}\rangle$, so the probability per path state is of the same order for both.

\section{Conclusion}
We have found that in the case of chains of two and three stars with multiple paths between two distinguished vertices, a quantum walk can find  those paths with a quantum speedup..  The case of two stars is straightforward, and there is only a single frequency in the problem that tells us when to measure the position of the particle.  In the case of three stars, there are two frequencies, but the dynamics of the state of the walk is largely determined by one of them, so we again know when to measure the particle in order to find a path state.  Previous examples of using a quantum walk to find a path took place on trees, that is graphs with no cycles.   The graphs considered here do have cycles, so we see that path finding by means of a quantum walk is not confined to tree graphs.  These examples indicate that quantum walks can find a way through a maze when is more than one possible path.

\end{document}